\documentclass[conference]{IEEEtran}
\usepackage[T1]{fontenc}
\IEEEoverridecommandlockouts
\usepackage{cite}
\usepackage{amsmath,amssymb,amsfonts}
\usepackage{algorithmic}
\usepackage{graphicx}
\usepackage{textcomp}
\usepackage{afterpage}
\usepackage[hyphens]{url}
\usepackage{hyperref}
\usepackage{xcolor}
\usepackage{hyperref}
\usepackage{amsmath}    
\usepackage{amssymb}    
\usepackage{amsthm}     
\usepackage{bm}         
\usepackage{booktabs}

\usepackage{tikz}
\usepackage{pgfplots}
\usepackage{caption}
\usetikzlibrary{shapes.geometric, arrows, positioning}

\usepackage{hyperref}   
\hypersetup{
    colorlinks=true,
    linkcolor=blue,
    citecolor=blue,
    urlcolor=blue
}

\usepackage{tabularx}
\usepackage{caption}    
\usepackage{subcaption} 
\usepackage{multirow}   

\usepackage{siunitx}    
\def\BibTeX{{\rm B\kern-.05em{\sc i\kern-.025em b}\kern-.08em
    T\kern-.1667em\lower.7ex\hbox{E}\kern-.125emX}}

\usepackage{hyperref}   
\hypersetup{
    colorlinks=true,
    linkcolor=blue,
    citecolor=blue,
    urlcolor=blue
}

\usepackage{tabularx}
\usepackage{caption}    
\usepackage{subcaption} 
\usepackage{multirow}   

\usepackage{siunitx}    
\def\BibTeX{{\rm B\kern-.05em{\sc i\kern-.025em b}\kern-.08em
    T\kern-.1667em\lower.7ex\hbox{E}\kern-.125emX}}
\begin{document}

\title{Smart Low-Carbon Freight Transport:\\
A Comparative Analysis}
\author{

\IEEEauthorblockN{Chiamaka Anicho-Okoro\IEEEauthorrefmark{1}, Mariem Belhor\IEEEauthorrefmark{2} and Omar Alam\IEEEauthorrefmark{1}}
\IEEEauthorblockA{
\IEEEauthorrefmark{1}Trent University, Peterborough, Ontario, Canada\\
\IEEEauthorrefmark{2} University of Picardie Jules Verne, Laboratoire des Technologies Innovantes (LTI) UR 3899, Amiens, France\\
Emails: chiamakaanichookoro@trentu.ca, mariem.belhor@u-picardie.fr, 
omaralam@trentu.ca}
}

\maketitle

\begin{abstract}
The decarbonization of freight transport is a major challenge, particularly with the growing demand for last-mile delivery. This paper focuses on low-carbon transport modes for last-mile logistics and presents a comparative analysis based on environmental and operational performance indicators. The evaluation highlights differences in emissions across delivery modes and the impact of factors such as distance and traffic conditions. In addition, the role of emerging technologies is discussed, emphasizing how IoT and AI can support real-time monitoring and decision-making. The results underline the importance of combining low-carbon transport solutions with data-driven approaches to improve both efficiency and sustainability in freight operations.

\end{abstract}



\section{Introduction}
Freight transport is a major component of transportation systems and a significant source of greenhouse gas emissions. In the European Union, the transport sector accounts for approximately 27\% of total greenhouse gas emissions~\footnote{Eurostat, \textit{Key Figures on European Transport – 2025 Edition}. Available: \url{https://ec.europa.eu/eurostat/fr/web/products-key-figures/w/ks-01-25-057}}. Reducing the environmental impact of freight mobility therefore requires the development of low-carbon transport solutions supported by intelligent transportation systems capable of improving efficiency through data-driven and computational approaches~\cite{futuretransp5020034,OUYANG2025146906}.

Recent advances in information and communication technologies have enabled the deployment of smart freight transport systems based on real-time data acquisition and dynamic decision-making. Technologies such as the Internet of Things (IoT) facilitate the monitoring of transport operations~\cite{OUEDRAOGO2026113108, WangDu2025ColdChainIoT }, while artificial intelligence (AI) enables the processing of large-scale data for optimization \cite{drones9030158,  belhor2023fuzzy}, prediction \cite{belhor2024enhanced, PEGADOBARDAYO2024110665}, and system control \cite{belhor2025drone, CORDEIRO2025100038}. In addition, autonomous systems, including drones, represent emerging transport modes that rely heavily on embedded intelligence and real-time coordination \cite{belhor2025drone}.

In this context, the selection of an appropriate freight transport mode can no longer rely only on traditional economic or environmental criteria. It must also integrate computational and data-driven dimensions, including optimization, predictive modeling, and the ability to support real-time decision-making. 
This paper addresses this issue by proposing a comparative evaluation of freight transport modes based on multiple indicators, including carbon emissions.
\begin{figure*}[t]
    \centering
    \fbox{\includegraphics[width=.9\textwidth]{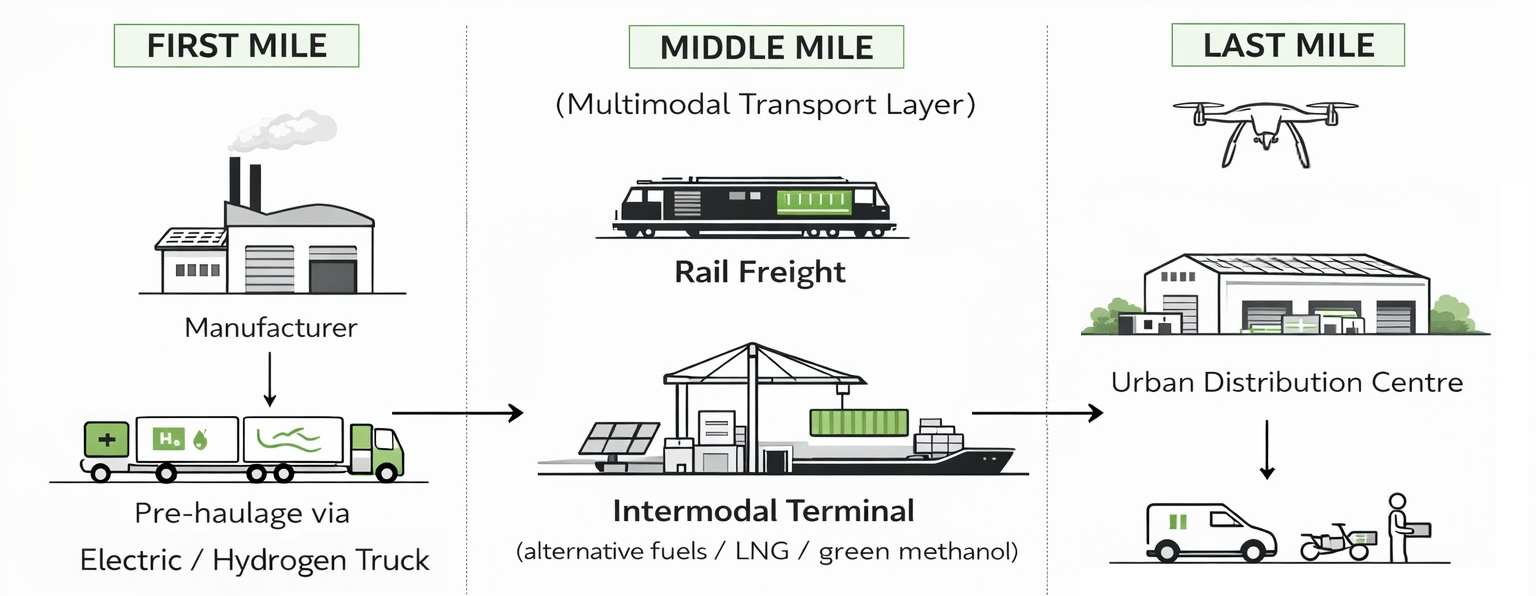}}
    \caption{Multimodal Low-Carbon Freight Distribution Process}
    \label{fig:multimodal}
\end{figure*}
The remainder of this paper is organized as follows. Section~\ref{sec:section2} presents an overview of eco-friendly and low-carbon freight transport solutions. Section~\ref{sec:section3} introduces emerging intelligent technologies for freight transport. Section~\ref{sec:section4} reports the comparative analysis and discusses the results. Section~\ref{sec:conclusion} concludes the paper and outlines future research directions.

\section{Eco-Friendly and Low-Carbon Freight Transport Solutions}
\label{sec:section2}

To cover representative low-carbon options across freight transport, we consider transport modes operating across different logistics layers. While some modes are directly involved in last-mile delivery, others contribute through multimodal freight systems that enable low-carbon urban distribution.

a) Electric road freight: Electric Vehicles (EVs) represent one of the primary solutions for last-mile delivery. They are widely deployed in urban logistics due to their operational flexibility and zero tailpipe emissions. In recent years, a growing body of research has focused on the optimization and large-scale deployment of electric delivery fleets \cite{YANG2024104127,MOON2024100351, DIMATULAC2025104059}, addressing challenges such as routing under battery constraints \cite{ERDELIC2026111758,app15094703}, charging infrastructure planning  \cite{DIMATULAC2025104059} and energy consumption modeling  \cite{WOO2024103644,SHEIKHASADI20252321}.

b) Rail freight: Rail transport is primarily used for long-haul freight movement and is not directly involved in last-mile delivery. However, within multimodal logistics chains, rail systems supply urban consolidation centers and intermodal terminals that support last-mile distribution networks~\cite{JIANG2025137068,GILLSTROM2024972}.

c) Maritime and inland waterways: Waterborne freight transport operates primarily in upstream logistics for bulk and containerized flows. In multimodal configurations, goods transported via maritime shipping or inland waterways are transferred to low-carbon urban fleets for final delivery~\cite{Abu-Aisha2024SeaRailSLR,SahaKhalil2025DryPorts,en18195146}.

d) Urban micro-mobility freight: Cargo bikes and e-cargo bikes are specifically designed for last-mile delivery in dense urban environments \cite{GALKIN2025104609, futuretransp5010031}. They offer high maneuverability and near-zero emissions.

e) Multimodal solutions: Integrated logistics systems combining rail or waterways with green vehicles (e.g. electric vans) enable low-carbon freight distribution, where the final delivery segment is performed by zero-emission modes. In these multimodal architectures, the freight is first transported over long distances using energy-efficient infrastructures such as electrified rail networks or inland waterways. Goods are then transferred at urban consolidation centers or intermodal terminals located at the periphery of cities, where they are consolidated and dispatched for last-mile delivery~\cite{GILLSTROM2024972,SahaKhalil2025DryPorts}.

From these hubs, last-mile delivery is carried out using electric vans or low-emission hybrid vehicles for suburban and medium-distance distribution (typically short-to-medium distances from the urban distribution center) \cite{YANG2024104127}. In dense urban areas, where traffic restrictions, low-emission zones, and limited parking constrain conventional vehicle access \cite{ALVAREZHORCAJO2025100282}, cargo bikes and electrically assisted cargo bikes \cite{KlatteKuhnimhof2025CargoBikesDoubleParking} are increasingly used due to their operational flexibility and minimal space requirements.

More recently, in countries such as the United States, projects have introduced delivery drones for small parcels\footnote{Walmart Takes Flight With Drone Delivery Expansion to Five New Cities, Redefining Fast, Flexible Retail, Jun. 5, 2025.\\
\url{https://corporate.walmart.com/news/2025/06/05/walmart-takes-flight-with-drone-delivery-expansion-to-5-new-cities-redefining-fast-flexible-retail}}, particularly in suburban or peri-urban areas where road congestion or geographical constraints limit the efficiency of conventional vehicles \cite{drones9110759, drones9030158}.

As illustrated in Fig.~\ref{fig:multimodal}, such multimodal configurations generate several operational and environmental advantages:

\begin{itemize}
\item First, they reduce urban road congestion by shifting long-haul freight flows to rail or maritime transport before entering metropolitan areas.
\item Second, hub-based transshipment improves load consolidation rates and vehicle utilization, particularly when shipments from multiple origins are aggregated at the intermodal terminal.
\item Third, the progressive substitution of conventional diesel trucks with EVs and, increasingly, hydrogen fuel-cell trucks for medium- and long-haul segments significantly reduces greenhouse gas emissions across the logistics chain.
\end{itemize}
Recent research highlights the growing interest in hydrogen-powered freight fleets, particularly for applications in which EV solutions face limitations related to range, payload, or charging time \cite{wevj16020076, TIAN2025135797}. Hydrogen fuel-cell vehicles can refuel quickly and provide long driving ranges, making them attractive for regional and heavy-duty freight operations. However, infrastructure constraints and high capital costs remain significant barriers to large-scale deployment.
From a system coordination perspective, the architecture shown in Fig.~\ref{fig:multimodal} requires synchronization between upstream transport schedules (rail or maritime arrivals) and downstream last-mile routing operations. This coordination is typically enabled by digital freight management platforms, real-time tracking systems, and optimization algorithms that support dynamic routing and efficient cross-docking at the urban distribution center.

The reviewed transport modes highlight that no single mode dominates across all objectives. Electric vans provide practical decarbonization for last-mile delivery but depend on charging infrastructure and electricity carbon intensity. Rail and waterways offer strong emission performance for long-haul transport but require intermodal integration for last-mile distribution. Cargo bikes are highly effective in dense urban environments yet remain constrained by payload and range. Emerging solutions such as drones can address specific last-mile scenarios but introduce new constraints related to payload capacity, safety, and regulatory requirements.

Coordinating these diverse transport modes requires balancing service reliability, cost efficiency and environmental performance within a unified decision-making framework. These coordination challenges highlight a main research question of this work:

\emph{How can intelligent and data-driven technologies support the integration and optimization of multi-modal low-carbon freight systems while addressing operational, infrastructural and environmental constraints?}

\section{Emerging Smart Technologies for Freight Transport}
\label{sec:section3}

The digital transformation of freight transport is driven by the rapid development of smart technologies that enable real-time monitoring, intelligent decision-making and system-wide optimization \cite{mohsen2024ai}. This section presents the main emerging technologies supporting the transition toward intelligent and low-carbon freight transport systems.

\subsection{IoT for Real-Time Freight Monitoring}

\begin{figure*}
    \centering
    \fbox{\includegraphics[width=.5\linewidth]{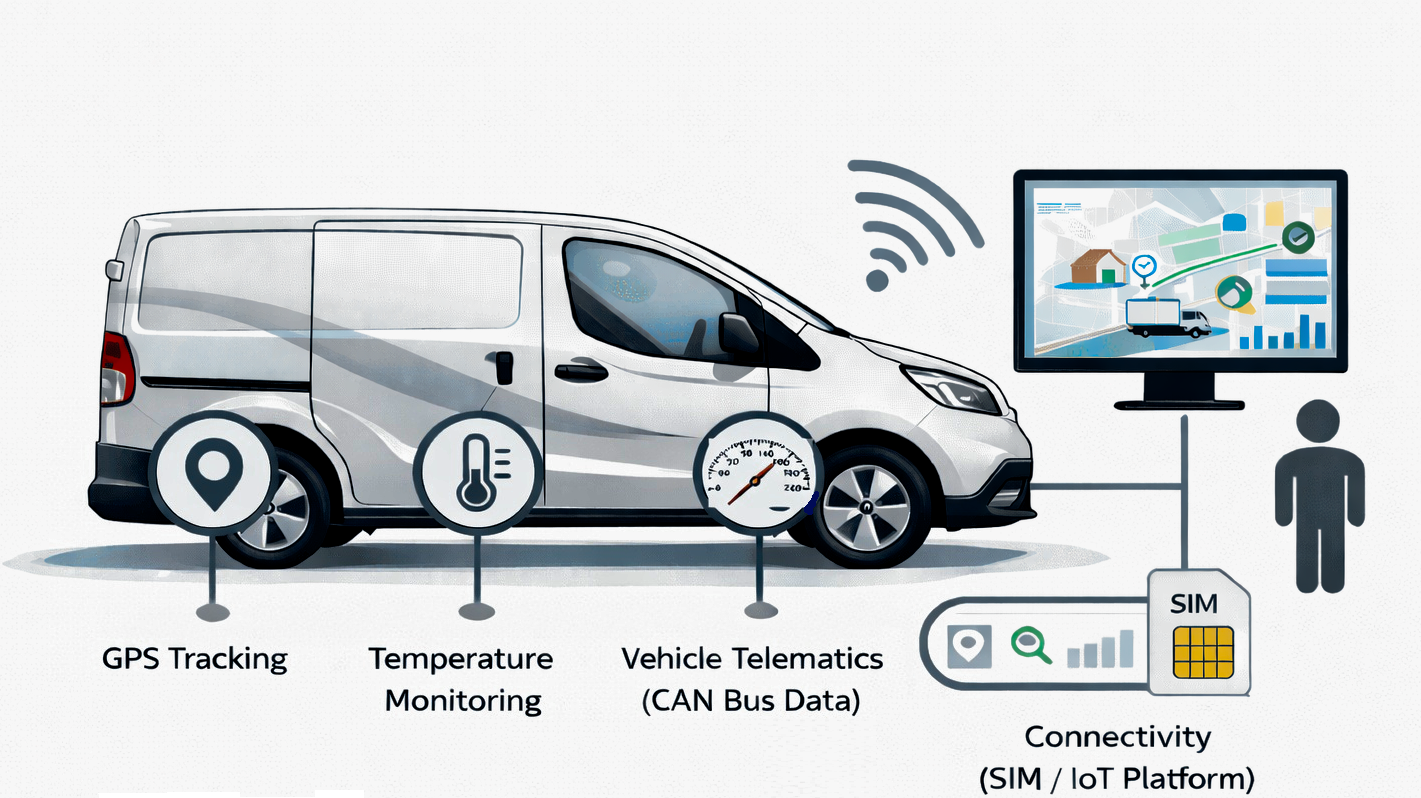}}
    \caption{IoT-enabled freight vehicle monitoring system}
    \label{fig:iot}
\end{figure*}

The IoT enables real-time visibility across freight transport operations. Connected sensors and tracking devices continuously collect data on vehicle location, cargo status, environmental conditions, and operational performance \cite{huynh2024iot}, as illustrated in Figure~\ref{fig:iot}. The figure presents the main components of an IoT-based freight transport system, including GPS tracking, temperature monitoring, vehicle telematics (CAN bus data), and connectivity through a SIM-enabled IoT platform. These technologies support real-time data collection, enhance operational visibility, and enable data-driven decision-making for more efficient and low-carbon freight transport management.

AI techniques are increasingly used to process large-scale transport data and support optimization tasks \cite{belhor2020new,danach2025neuhh}. Machine learning and predictive analytics enable demand forecasting, travel time estimation and route optimization under dynamic conditions. AI-driven decision systems can continuously adjust transport plans based on real-time inputs, improving operational efficiency while reducing fuel consumption and emissions. Such approaches are particularly relevant for last-mile delivery, where uncertainty and variability are high.

\subsection{Digital Twins and Blockchain for Intelligent Freight Systems}

\begin{figure*}[h]
    \centering
    \includegraphics[width=.6\linewidth]{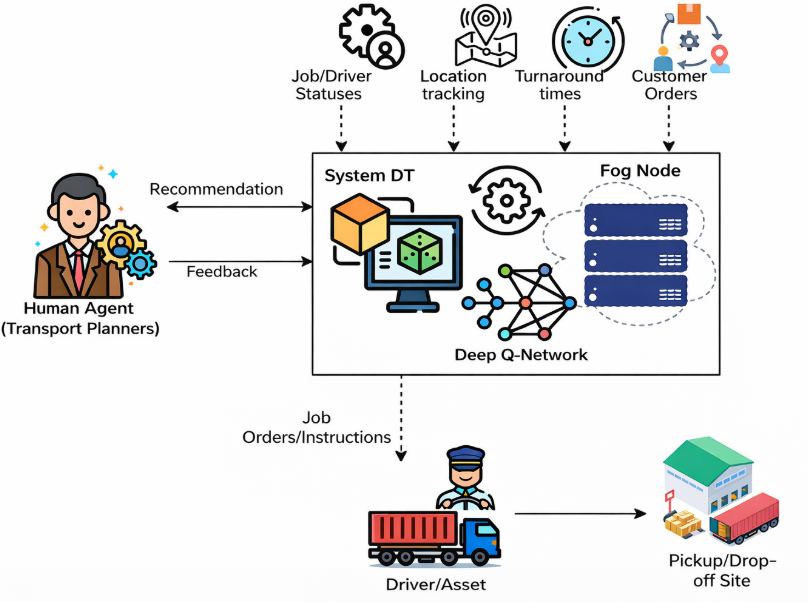}
    \caption{Digital Twin-Based Intelligent Freight Transport System}
    \label{fig:digital}
\end{figure*}

Digital twin technology enables the creation of virtual replicas of freight transport systems for real-time simulation and performance analysis. These models support scenario testing, infrastructure planning, and operational optimization before physical deployment. In parallel, blockchain technologies enable secure and transparent data exchange among logistics stakeholders, improving trust, traceability, and data integrity in connected freight systems \cite{mullet2026simulation}, \cite{karaduman2025blockchain}. Figure~\ref{fig:digital} illustrates a digital twin-based freight transport system that integrates real-time data such as driver status, location tracking, turnaround times, and customer orders. The digital twin interacts with human planners by providing recommendations and receiving feedback, while optimizing operations and generating instructions for drivers and assets.

These emerging technologies constitute together the digital backbone of next-generation freight transport systems and are particularly relevant for improving the performance indicators analyzed in Section~\ref{sec:section4}, including emission efficiency and congestion sensitivity.
\subsection{AI for Freight Transport Optimization}
\section{Comparative Analysis and Results}
\label{sec:section4}

This section presents a comparative evaluation of freight transport modes using a multi-criteria analysis. The objective is to assess the performance of conventional and emerging last-mile delivery solutions with respect to environmental and operational indicators. The analysis is statistical and based on aggregated delivery observations derived from a public dataset.

\subsection{Dataset Description}

The analysis is based on a publicly available dataset obtained from Kaggle\footnote{Green Logistics Carbon Footprint for E-commerce Dataset: \url{https://www.kaggle.com/datasets/bertnardomariouskono/green-logistics-carbon-footprint-for-e-commerce}}. The dataset contains approximately 12,000 delivery records and is designed to support the environmental assessment of urban freight transport systems.

Each record corresponds to a delivery operation and includes key attributes such as delivery distance, package weight, transport mode, and traffic congestion level. Carbon emissions are expressed in kilograms of CO$_2$ equivalent (kgCO$_2$e) and are estimated using a physics-based modeling approach based on emission factors.

More specifically, each vehicle category is associated with an average emission coefficient (in gCO$_2$/km), which is combined with delivery distance and operational conditions to estimate total emissions per trip. For instance, diesel vans typically emit around 200 gCO$_2$/km, while EVs generate approximately 50 gCO$_2$/km when accounting for electricity generation.

\subsection{Evaluation Criteria}

To evaluate the sustainability and operational performance of delivery modes, a set of key performance indicators (KPIs) is defined.

Let $\mathcal{M}$ denote the set of delivery modes and $m \in \mathcal{M}$ a given mode. Let $\mathcal{D}_m$ be the set of deliveries performed using mode $m$, with $N_m = |\mathcal{D}_m|$ the number of deliveries. Each delivery $i \in \mathcal{D}_m$ is associated with:
\begin{itemize}
    \item $E_i$: carbon emissions (kgCO$_2$e)
    \item $d_i$: delivery distance (km)
\end{itemize}

The following KPIs are computed:

\textbf{Average Emission per Delivery}
\begin{equation}
\mathrm{AvgCO_2}(m) = \frac{1}{N_m} \sum_{i \in \mathcal{D}_m} E_i
\end{equation}

\textbf{Emission Intensity per Kilometer}
\begin{equation}
\mathrm{CO_2/km}(m) = 
\frac{\sum_{i \in \mathcal{D}_m} E_i}{\sum_{i \in \mathcal{D}_m} d_i}
\end{equation}

\textbf{Congestion Sensitivity Ratio (CSR)}
\begin{equation}
\mathrm{CSR}(m) =
\frac{\mathrm{AvgCO_2}^{\text{severe}}(m)}
{\mathrm{AvgCO_2}^{\text{low}}(m)}
\end{equation}

where $\mathrm{AvgCO_2}^{\text{severe}}(m)$ and $\mathrm{AvgCO_2}^{\text{low}}(m)$ denote the average emissions computed over deliveries under severe and low congestion conditions, respectively.
Table~\ref{tab:kpi_results} summarizes the computed KPIs for each delivery mode. The columns report the average carbon emissions per delivery (in kgCO$_2$e), the emission intensity per kilometer (kg/km), and the CSR, which measures the relative increase in emissions under severe traffic conditions compared to low congestion levels. The CSR is not defined for cargo bikes due to their negligible operational emissions.
\begin{table}[h]
\centering
\caption{Measured KPI Results by Delivery Mode}
\label{tab:kpi_results}
\begin{tabular}{lccc}
\toprule
\textbf{Mode} & \textbf{Avg CO$_2$ (kg)} & \textbf{CO$_2$/km} & \textbf{CSR} \\
\midrule
Cargo Bike & 0.00 & 0.000 & -- \\
Drone & 0.06 & 0.008 & 1.09 \\
Electric Van & 4.06 & 0.051 & 1.20 \\
Motorcycle & 2.51 & 0.112 & 1.86 \\
Diesel Van (Euro 6) & 17.12 & 0.214 & 1.76 \\
Diesel Van (Euro 4) & 25.91 & 0.317 & 1.75 \\
Heavy Truck & 463.14 & 1.036 & 1.70 \\
\bottomrule
\end{tabular}
\end{table}
\subsection{Discussion}

The results reveal substantial environmental disparities across last-mile transport modes. Electric vans reduce average emissions by approximately 84\% compared to Euro 4 diesel vans (4.06 vs 25.91 kgCO$_2$ per delivery), while maintaining comparable delivery distances. 
Micro-mobility solutions such as cargo bikes and drones exhibit near-zero operational emissions, making them highly suitable for short-distance urban delivery. However, their applicability remains limited to small payloads and short ranges.
Combustion-based vehicles show strong sensitivity to traffic congestion. Diesel fleets experience emission increases exceeding 70\% under severe congestion conditions (CSR $\approx$ 1.75), whereas electric vans demonstrate significantly lower sensitivity (CSR $\approx$ 1.20). 

\begin{figure}[h]
\centering
\includegraphics[width=0.48\textwidth]{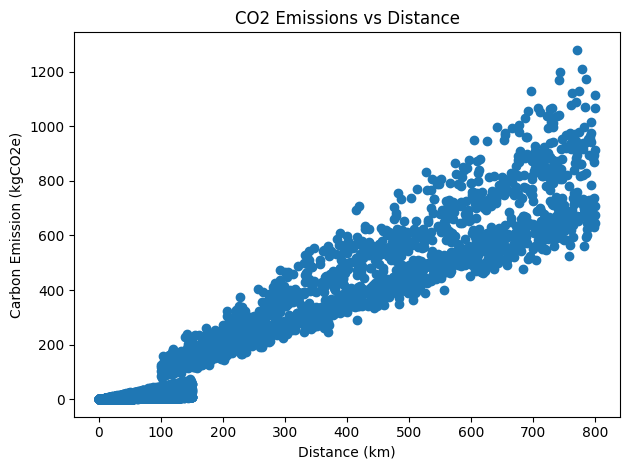}
\caption{Carbon emissions as a function of delivery distance}
\label{fig:distance}
\end{figure}
\begin{figure}[h]
\centering
\includegraphics[width=0.48\textwidth]{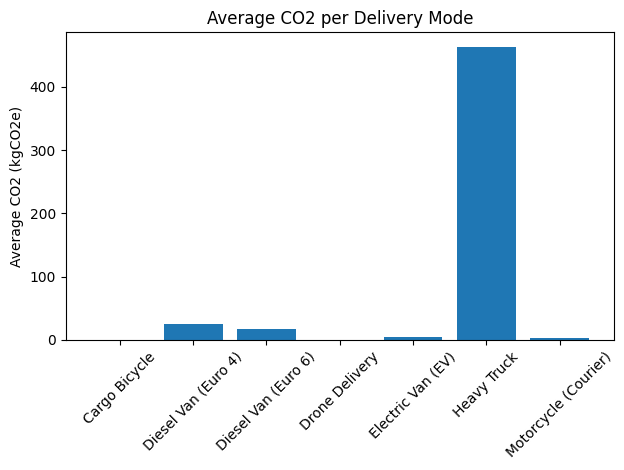}
 \caption{Average CO$_2$ emissions per delivery mode}
    \label{fig:avg_mode}
\end{figure}

Figure~\ref{fig:distance} shows that carbon emissions increase with delivery distance for all transport modes, with clear differences depending on the vehicle type. Figure~\ref{fig:avg_mode} confirms this by comparing average emissions across modes. The dispersion observed at longer distances reflects variations in payload and traffic conditions, which further affect energy consumption. The delivery mode strongly influences carbon intensity, while distance remains the main driver of emissions. Congestion and operational variability further impact performance, particularly for combustion-based vehicles. This highlights the importance of efficient routing and transport management.

 Electrified fleets significantly reduce emissions compared to diesel vehicles, while micro-mobility solutions achieve near-zero emissions for short-distance deliveries. No single transport mode optimizes all criteria. These results support the use of hybrid and multimodal strategies combining low-carbon solutions with data-driven optimization to improve both environmental and operational performance. In this context, emerging technologies such as IoT and AI enhance visibility, traceability and decision-making, contributing to reduced emissions, improved efficiency, and more reliable freight operations.

\section{Conclusion}
\label{sec:conclusion}

This paper presented a comparative analysis of low-carbon transport modes for last-mile delivery based on environmental performance indicators.

The results show that EVs significantly reduce emissions compared to conventional diesel fleets, while micro-mobility solutions such as cargo bikes and drones achieve near-zero emissions for short-distance deliveries. However, their applicability remains limited by payload, range, and infrastructure constraints. No single transport mode optimizes all criteria.

The analysis also highlights the impact of operational factors such as delivery distance and traffic congestion, with combustion-based vehicles showing higher sensitivity compared to electric alternatives.

These results support the need for hybrid and multimodal last-mile delivery strategies combining complementary transport modes. The integration of data-driven technologies further enhances operational efficiency and emission reduction.

Future work will extend the analysis using larger datasets and develop optimization models for dynamic last-mile transport mode selection under real-world constraints.


\bibliographystyle{IEEEtran}
\bibliography{References}

\end{document}